\documentclass[12pt]{article}
\usepackage{epsfig}

\begin{document}

\begin{center}
{\bf
INVESTIGATION OF ZERO-SOUND DISPERSION EQUATION IN THE COMPLEX PLANE OF
FREQUENCY
}

\vspace{1cm}

 V.A. Sadovnikova\\
Petersburg Nuclear Physics Institute, \\
Gatchina, St.~Petersburg 188300,
Russia

\vspace{1cm}

{\bf Abstract}

The known solutions to the zero-sound dispersion equation are
considered as placed on the physical and unphysical sheets
 in the complex plane of  frequency.
\end{center}

\begin{center}
{\bf I. Introduction}
\end{center}
In this paper we present the known solutions to the
 zero-sound dispersion equation obtained in the Landau  kinetic
theory  \cite{La}-\cite{PN} and in the microscopic
theory in RPA \cite{PN}-\cite{FW} in the complex plane of
frequency. The solutions are calculated for the different values
of coupling constant and considered on the physical and
unphysical sheets.

In recent years a large attention is paid to the problem of
stability of the nuclear matter and the phase transitions in it
\cite{Mi}, \cite{PR}, \cite{MS}. To investigate the phase
transitions, it is useful to understand what kind of excitations
becomes amplified and leads to unstability of matter with the
changing of the temperature or density or coupling constant. In
the fermi-liquid the amplified solutions appear at $F_0<-1$.
When we take into account the meson exchange then the zero-sound
modes undergo the drastic modifications and the damping
solutions can  turn into the amplified ones and lead to the appearance
of the unstability of the description of nuclear matter.

In this note the solutions to the  zero-sound dispersion equation are
presented. They are given in such a form that permits to trace
the influence of the external field and meson exchange on these
solutions later on.

Following the papers \cite{LaLi}, \cite{FW} we write the dispersion
equation in the form
\begin{equation}\label{1}
1 =  {\cal F} \Pi^{0R}(\omega,k).
\end{equation}
In this equation ${\cal F}$ is the effective quasiparticle
interactions, $\omega$ and $k$ are the frequency and the wave
vector of the excitations. The operator $\Pi^{0R}(\omega,k)$  is
the retarded zero order on the quasiparticle interaction
self-energy part (polarization operator).  There is a known
analytical expression for this polarization operator
\cite{LaLi}-\cite{FW}.

It was shown in papers  \cite{LaLi}, \cite{Mi}, that the energies
of the zero-sound collective excitations are the poles of the
two-quasiparticle Green function over the
quasiparticle-quasihole channel.  The two-quasiparticle Green
function $G^{(2)}$ is determined through the
single-quasiparticle Green function $G^{(1)}$ and the vertex
function $\Gamma$

$$
G^{(2)} = G^{(1)}G^{(1)} + G^{(1)}G^{(1)}\Gamma G^{(1)}G^{(1)} .
$$

When we consider the interacting quasiparticles near the  Fermi
surface with the small momenta transmitted then the vertex
function $\Gamma$ can be given by \cite{Mi}
\begin{equation}\label{m1}
\Gamma(\omega,k) =\frac 1{a^2}\frac{{\cal F}}{1-{\cal
F}\Pi^{0R}(\omega,k)} ,
\end{equation}
where $a$ is a residue in the pole of $G^{(1)}$. In this equation the
constant interaction between the quasiparticles is supported. On the
other hand, $\Gamma$ is the two-particle scattering amplitude that
satisfy to the two-particle equation like Bethe-Salpeter one for
the particle and hole \cite{LaLi}.

The analytical singularities of the two-particle scattering
amplitude in the  complex plane of frequency are well known
\cite{Tlr}. On the real axis of the physical sheet of $\omega$
the stable excitations and cuts are placed . The cuts correspond
to the excitation of the free  particle-hole pairs. The damping
excitations (resonances) are disposed on the unphysical sheets under
the cuts. Besides, in the fermi-liquid at a strong attraction between
quasiparticles $F_0< -1$ the imaginary solutions to the Eq.(\ref{1})
appear. They are situated on the positive and negative imaginary axes
of the physical sheet. These solutions break the causality conditions
and point out the phase transition in nuclear matter. In the
present paper we consider the stable, damping and increasing solutions
of Eq.(\ref{1}) for the different magnitudes of the quasiparticle
interaction.

Following to \cite{PN}, \cite{Tlr} we determine the physical
sheet of the complex plane of frequency on which the stable
solutions of Eq.(\ref{1}) are situated.  At some values of
${\cal F}$ the solutions are placed both on physical and
unphysical sheets. These unphysical sheets belong to the
frequency Riemann surfaces that are determined by the form of
$\Pi^{0R}(\omega,k)$.

 We take ${\cal F}$ as a constant
scalar interaction ${\cal F}=F_0C_0$.  Here $C_0$ is  a factor
 inversed to the level density on the Fermi surface for  two
kinds of nucleons, $C_0=\pi^2/(2mp_F)$. $F_0$ is a dimensionless
coupling constant of the scalar quasiparticle interaction.

We consider Eq.(\ref{1}) in the kinetic theory of Landau
\cite{LaLi}-\cite{PN} and in the microscopic theory in the
random phase approximation (RPA) \cite{LaLi}-\cite{FW}. The
results of these theories coincide at the long wavelenths. For
the fixed $F_0$  and Fermi momenta $p_F$, the solutions to
Eq.(\ref{1}) are the linear functions of the frequency $\omega$
on the wave vector $k$:  $\omega =\ k s\frac{p_F}m$ in the
kinetic theory.  In RPA there is a more composite dependence, and
we have  branches of solutions $\omega(k)$.

In Sect.2 the analytical continuation of $\Pi^{0R}(\omega,k)$ on
the unphysical sheets of the complex plane is realized.

In Sect.3  the solutions of Eq.(\ref{1}) for the attractive
interaction $F_0 < 0$ are presented. It is shown that the part
of solutions to (\ref{1}) is situated on the unphysical sheets
(sheets $I$ and $I'$).

In Sect.4  the solutions for the repulsive interaction $F_0 > 0$
are presented. It is shown how after the overlapping of the
frequency of the zero-sound excitations, $\omega_s(k)$, and the
particle-hole continuum,  the branch $\omega_s(k)$ goes to the
unphysical sheet (sheets  $II$ and $II'$).

In the paper   we consider the symmetric nuclear matter at zero
temperature and the equilibrium density $\rho=\rho_0$,
$p_F=0.268$~GeV. During the  computations the mass of the
quasiparticles is taken equal to $m=0.8m_0$ and $m_0=0.94$~GeV.

\begin{center}
{\bf II. The structure of polarization operator}
\end{center}

In this section the expression for a polarization operator is
given. The formulae repeat the known expressions \cite{FW},
\cite{PN}. But they are presented in a form that is more
convenient for the analytical continuation of
$\Pi^{0R}(\omega,k)$ in $\omega$ from the physical on unphysical
sheets. At the beginning we consider the expression for the
polarization operator in RPA.

Recall that there is a condition on the magnitude of the
retarded operator $\Pi^{0R}(\omega,k)$ in the  complex plane of
$\omega$ \cite{LaLi}
\begin{equation}\label{2}
\Pi^{0R}(-\omega^*) = (\Pi^{0R})^*(\omega) .
\end{equation}
The expression for $\Pi^0$  is taken from \cite{PN}-\cite{FW}.
Let us  write it in the form that does not content the
overlapping logarithmic cuts. The causal operator $\Pi^0$ can be
presented as a sum of two terms. One term describes the
excitation  of the  particle-hole pair in the medium and the
second one corresponds to absorption of it
\begin{equation}\label{3}
\Pi^0(\omega,k) = \phi(\omega,k) + \phi(-\omega,k).
\end{equation}
For the real $\omega$, the retarded polarization operator
is determined by the following way \cite{FW}
$$ \Pi^{0R}(\omega,k) = (Re + i\ sign(\omega) Im)
\Pi^0(\omega,k).
$$
The expression for $\phi(\omega,k)$ is written for the
excitations in nuclear matter consisting of two sorts of
nucleons with two spin projection.

For the wave vectors $0\le k\le 2p_F$  the term $\phi(\omega,k)$
has a form
\begin{equation}\label{10}
\phi(\omega,k)=
-4\frac{m}{k} \frac1{4\pi^2}
\left(
\frac{-\omega m+kp_F}2-\omega m \ln\left(\frac{\omega m}{\omega
m-kp_F+\frac12k^2}\right)\right.
\end{equation}
$$
+\left.
\frac{(kp_F)^2-(\omega m-\frac12k^2)^2}{2k^2}
\ln\left( \frac{\omega
m-kp_F-\frac12k^2}{\omega m-kp_F+\frac12k^2}\right)\right) .
$$
For $k\ge2p_F$ the term $\phi(\omega,k)$ is a Migdal's function
\cite{Mi}:
\begin{equation} \label{11}
\phi(\omega,k) =\ -4\frac1{4\pi^2}\
\frac{m^3}{k^3}
\left[\frac{a^2-b^2}2 \ln\left(\frac{a+b}{a-b}\right)-ab\right]\
,
\end{equation}
where $a=\omega-(\frac{k^2}{2m})$, $b=\frac{kp_F}{m}$.

Consider the cuts of $\Pi^0(\omega,k)$ in the complex plane of
$\omega$. From (\ref{10}) we see that there are two cuts for
$k\le2p_F$ (denote them  $I$ and $II$). They are determined by
the first and the second logarithms in (\ref{10}) and are placed
on the real axis at $\omega$  equal to
\begin{equation} \label{12}
I: 0 < \omega\ <
\frac{kp_F}{m} -\frac{k^2}{2m}\, \qquad
 II:
\frac{kp_F}{m}-\frac{k^2}{2m}\ < \omega\ <
\frac{kp_F}{m}+\frac{k^2}{2m}\ .
\end{equation}
It is easy to see that $\phi(\omega,k)$ (\ref{10}) is finite in
the branch point $\omega m=kp_F-\frac12k^2$ due to the
cancelation of the infinite contributions of the first and the
second logarithms.

The cuts $I$ and $II$ can be considered as corresponding to the
 excitations of the different particle-hole pairs.
The cut $I$ describes
the excitation of a hole with the energy
$\varepsilon_{\vec{p}_F-{\vec k}}$ and a particle on the
Fermi surface. Then $\omega = \frac{p^2_F}{2m} -
\frac{(\vec{p}_F-{\vec k})^2}{2m}$ and the points of the cut
correspond to the change of the angle  $\theta$ between
$\vec{p}_F$ and $\vec k$ in the interval $\frac
k{2p_F}<cos\theta<1$.  The cut $II$ describes the excitation of
a particle with the energy $\varepsilon_{\vec{p}_F+{\vec k}}$
and a hole on the Fermi surface. Then
$\omega = \frac{(\vec{p}_F+{\vec k})^2}{2m} - \frac{p^2_F}{2m}$
and the points of the cut
correspond to $1-\frac k{p_F}<cos\theta<1$.

The cuts of the function $\phi(-\omega,k)$  lie on the negative
real axis symmetrically with respect to the cuts of
$\phi(\omega,k)$.  Thus, $\Pi^0(\omega,k)$ has four cuts in the
complex plane of $\omega$ which are shown in Fig.1a. When
$\omega>0$ then the cuts stem from $\phi(\omega,k)$ while for
$\omega<0$ they are caused by $\phi(-\omega,k)$.

For the both of the logarithms in (\ref{10}) the
infinite-sheeted Riemann surface can be construct
\cite{LSh}. The logarithms are the single-valued
function in it.

When $k$ grows, approaching $2p_F$, cut $I$ becomes shorter and
degenerates into a point at $k= 2p_F$. For $k>2p_F$ there are
 two cuts on the real axis in Eq.(\ref{3}): $III$ and symmetrical
$III'$.
\begin{equation}\label{12a}
III: -\frac{kp_F}{m} + \frac{k^2}{2m} < \omega <
\frac{kp_F}{m} + \frac{k^2}{2m}\ .
\end{equation}

Now we define the physical sheet according to the papers
\cite{LaLi}, \cite{PN}. In the long wavelenths
limit $k\rightarrow 0$, $\frac{\omega}{k}\rightarrow const$,
$\Pi^0(\omega,k)$ (\ref{3}) has a form
\begin{equation}\label{4}
\Pi^0(\omega,k)\approx
-4\frac{m}k\ \frac1{4\pi^2}
\left(\frac{-\omega m+kp_F}2
-\omega m \ln\left(\frac{\omega m}{\omega m-kp_F}\right)
\right.
\end{equation}
$$
+ \left.\frac{\omega m+kp_F}2-
\omega m \ln\left(\frac{-\omega m-kp_F}{-\omega m}\right)+ kp_F
\right)
$$
\begin{equation}\label{4a}
= -2 \frac{m p_F}{\pi^2}\left(1-
\frac{\omega m}{2 k p_F}
\ln
\frac{\frac{\omega m}{ k p_F} +1}{\frac{\omega m}{ k p_F}-1}
\right) .
\end{equation}
We accept that on the physical sheet the logarithm in (\ref{4a})
has  an imaginary part on the upper edge of the cut
equal to $-\pi i$ \cite{PN} and, consequently,  $+\pi i$  on the  lower
edge.

The corresponding retarded $\Pi^{0R}$ is (we denote
$s=\frac{\omega m}{k p_F}$)
\begin{equation}\label{4b}
\Pi^{0R}= 2 \frac{m p_F}{\pi^2}
\left(-1 + \frac s2\left[ ln|\frac{s+1}{s-1}|
-i\pi\theta(1-|s|)\right]\right) .
\end{equation}

Substituting Eq.(\ref{4a}) in (\ref{1}) we obtain the known
form of the zero-sound dispersion equation in the kinetic theory
\begin{equation}\label{5}
1+\frac 1{F_0} = \frac{\omega m}{2kp_F}ln\left(\frac
{\omega m + kp_F}{\omega m - kp_F}\right) =
\frac{s}{2}\ln\frac{s+1}{s-1} .
\end{equation}
In Eq.(\ref{4b}) there is a cut on the real axis for
 $(-\frac{kp_F}m < \omega <\frac{kp_F}m)$. Looking at
Eq.(\ref{4}) we  consider this cut as consisting of two cuts
$I$ and $I'$ (Fig.1a). At the long wavelenths Eq.(\ref{12}) comes
to $I: 0<\omega < \frac{kp_F}{m}$.

\begin{center}
{\bf III. Solutions at $F_0 < 0$}
\end{center}

In this section we consider solutions to
Eqs.(\ref{1}) and (\ref{5}) at $F_0<0$ in the complex plane of
$\omega$.

The family of solutions to Eqs.(\ref{1}) and (\ref{5}) at
$F_0<0$ we denote by $\omega_{sd}$. The letters "sd" are
connected with the words "spinodal decomposition". It is known
\cite{LaLi} that the thermodynamic stability condition
 says that the matter is stable if the partial derivative of the
pressure with respect to the volume is negative:
$\left(\frac{\partial P}{\partial V}\right)_T < 0$. The process
which takes place in matter, when this condition is broken, is
known as spinodal decomposition \cite{MS}. It is shown in
\cite{LaLi} that there is a relation of the effective
quasiparticle interaction to the partial derivative
$$
\frac{\partial P}{\partial V} = -\frac{N}{V^2} \frac{p_F^2}{3m}
(1 + F_0).
$$
($N$ is the number of particles).
The stability condition is broken at $F_0 < -1$. It looks
reasonable to use the notation $\omega_{sd}(k)$ for the
solutions at $F_0 < -1$. It is shown in this section that all
solutions at $F_0 < 0$ can be considered as belonging to the same
family since they continuously turn one into another at the
continuous change of $F_0$. Therefore the notation
$\omega_{sd}(k)$ is used for all solutions at $F_0< 0$.

\begin{center}
{\bf III.A The definition of the unphysical sheets}
\end{center}

Let us turn to Eqs.(\ref{3}) and (\ref{10}) and define the
unphysical sheets in the complex plane of $\omega$. Consider the
cut $I$.  We denote the first logarithm in (\ref{10}) as
$ln(z_1)\equiv ln\left(\frac{\omega m}
{\omega m-kp_F +\frac{k^2}2}\right)$.
While the frequency $\omega$ goes along the cut $I$ (\ref{12}):
$\omega=(0,\frac{kp_F}{m} -\frac{k^2}{2m})$ then $z_1$ is
changed in the interval $z_1=(0,-\infty)$. The cut $I$ in the
complex plane of $\omega$ corresponds to the  cut along the
negative real axis in the complex plane of $z_1$ (Fig.1b). When
we go to the cut $I$ from above (the arrow in Fig.1a),  this
corresponds that we go to the cut from below in the complex plane
of $z_1$ (the arrow in Fig.1b).
Going over under the cut in
Fig.1b we pass on to the unphysical sheet neighboring  with
the physical one on the Riemann surface of $ln(z_1)$. The
magnitudes of $ln(z_1)$ on the neighboring sheet differ by a quantity
of $(-2\pi i)$ from the magnitudes on the physical sheet at the same
$z_1$.

Thus, going over under the cut $I$ in Fig.1a we pass on to the
unphysical sheet of the complex plane of $\omega$  that is
placed under the physical sheet. Let us denote the unphysical
sheet with the same letter as a cut: $I$. The magnitude of
$ln\left(\frac{\omega m}{\omega m-kp_F +\frac{k^2}2}\right)$
on the sheet $I$ differ from the magnitudes on the physical sheet
by the quantity of $(-2\pi i)$.  As example, the value of
$ln\left(\frac{\omega m}{\omega m-kp_F +\frac{k^2}2}\right)$ on the
lower edge of cut $I$ is $ln\left(-\frac{\omega m}{\omega m-kp_F
+\frac{k^2}2}\right) +\pi i$. But the value at the same point  of
$\omega$ on the unphysical sheet is $ln\left(-\frac{\omega
m}{\omega m-kp_F +\frac{k^2}2}\right)-\pi i$.  Thus, we have the
continuous changing of logarithm along the arrow in Fig.1a. Then
we conclude that $\Pi^{0R}$ in  Eq.(\ref{1}) changes
continuously as well. The part of the unphysical sheet $I$ is
shown in Fig.1c by the shading with the right  slope.

An important remark is that moving on the frequency Riemann
surface of one of logarithms we stand on the physical sheet for
the other logarithms. In another words, the Riemann surfaces of
logarithms in Eq.(\ref{3}) are independent.

Now we define an unphysical sheet related to the cut $I'$ (denote
the sheet by $I'$). The cut $I'$ stems from the first logarithm
in $\phi(-\omega,k)$. We designate $ln(z_1')\equiv
-ln\left(\frac{-\omega m}
{-\omega m-kp_F +\frac{k^2}2}\right)$ = $
ln\left(\frac{\omega
m+kp_F -\frac{k^2}{2}} {\omega m}\right)$.
Going over under the cut of $ln(z_1')$ we add the shift $-2\pi
i$ to the magnitude of $ln(z_1')$. Then, the  logarithm
on the sheet $I'$  under the lower edge of the cut $I'$ is equal to
$ln\left(-\frac{\omega
m+kp_F -\frac{k^2}{2}} {\omega m}\right)+\pi i-2\pi i$.  This
expression is used for the analytical continuation to the sheet
$I'$. The part of the sheet $I'$ is shown on the Fig.1c by the
shading with the left slope.

The values of $\Pi^{0R}(\omega,k)$ on the unphysical
sheets $I$ and $I'$ defined by such a way are connected
by the relation
\begin{equation}\label{6a}
\left(\Pi^{0R}_I(\omega,k)\right)^* =
\Pi^{0R}_{I'}(-\omega^*,k) .
\end{equation}

Thus, we made two unphysical sheets $I$ and $I'$. Let us
show  that  the magnitudes of $\Pi^{0R}(\omega,k)$ (\ref{3})
on the negative imaginary axes of $I$ and $I'$ are real and
are the same,
according to (\ref{6a}). Then Eqs.(\ref{1}), (\ref{5}) and their
solutions coincide as well.

On the imaginary axis we denote $\omega=i\omega_i$.
On the sheet $I$ the operator $\Pi^{0R}(\omega,k)$ (\ref{3}) has
the following view on the negative imaginary axis
\begin{equation}\label{7}
\Pi^{0R}(i\omega_i,k) = -\frac {mp_F}{\pi^2}\left[1 -
\frac{i\omega_i m}{kp_F} \left(\ln\frac{-i\omega_i m}{i\omega_i
m-kp_F +\frac12 k^2} - \pi i\right) \right.
\nonumber
\end{equation}
$$ \left.
-\frac{i\omega_i m}{kp_F} \left(\ln\frac{i\omega_i m+kp_F
-\frac12 k^2}{i\omega_i m}\right) + \ ... \right].
$$
\begin{equation}\label{8}
= -\frac {mp_F}{\pi^2}
\left[1 +2\frac{\omega_i m}{kp_F}
    \left(arctg\frac{\omega_im}{kp_F-\frac12 k^2} - \frac\pi 2\right)
+  ...\right] \ .
\end{equation}

We leave only the first and the second terms in (\ref{10})
because only these terms have the additional shifts on the sheets
$I$ and $I'$. This expression is the analytical continuation of
$ln(-z_1) - \pi i$ to the imaginary axis.

On the negative imaginary axis of the sheet $I'$ we have for
$\Pi^{0R}(\omega,k)$
\begin{equation}\label{8a}
\Pi^{0R}(i\omega_i,k) = -\frac{mp_F}{\pi^2}
\left[1 -\frac{i\omega_i m}{kp_F}
\ln\frac{i\omega_i m}{i\omega_i m-kp_F +\frac12 k^2}
\right.
\nonumber
\end{equation}
$$ \left.
-\frac{i\omega_i m}{kp_F}
\left(\ln\frac{i\omega_i m+kp_F -\frac12 k^2}{-i\omega_i m}
    - \pi i\right) +  ... \right]
$$
and after the analogous transformation we obtain Eq.(\ref{8}) as
well.

We can conclude that there are unphysical sheets $I$ and $I'$,
that coincide on the imaginary axis. In the next section it is
shown that the solutions of Eqs.(\ref{1}), (\ref{5}) at $F_0 <0$
are disposed on imaginary axes both of the physical  and
 unphysical sheets.

\begin{center}
{\bf III.B Solution in the kinetic theory of Landau}
\end{center}

At $F_0<-1$ the positive and negative symmetrical solutions to
Eq.(\ref{5}) are placed on the imaginary axis of the physical
sheet (\cite{LaLi}, \cite{PN}). These solutions satisfy to the
equation  which can be deduced from Eq.(\ref{5})
($s=i\gamma$)
\begin{equation}\label{6}
1+\frac 1{F_0} = \gamma\ arctg\frac 1{\gamma}\ .
\end{equation}

The existence of the amplified solutions means that there is an
unstability of the nuclear matter.

Now we obtain the dispersion equation for the zero-sound
excitations at $-1<F_0<0$. We consider Eq.(\ref{8}) at long
wavelenths and then substitute it into Eq.(\ref{5}). Then on the
negative imaginary axis of the unphysical sheet we get an
equation
\begin{equation}\label{9}
1 + \frac 1{F_0} = -\gamma(arctg(\gamma) - \frac {\pi}{2}) .
\end{equation}
This equation was presented in \cite{CC} where it was mentioned
that there are no actual solutions of (\ref{9}). We really see
that the solutions to Eq.(\ref{9}) are placed on the unphysical
sheet ($I$).

In Fig.2  the solutions $\gamma$ of Eq.(\ref{5}) at $F_0<0$ are
presented. The curves in this figure are the same as the curves
in Fig.1 in papers \cite{CC} and \cite{PR} (except the line
$AB$).  The dashed domain means that the solutions are on the
unphysical sheets. Looking at Fig.2 we see how the overdamping
solutions  on the unphysical sheet turn into the growing
solutions and pass on to the physical sheet when the attraction
between quasiparticles increases.

Using the unphysical sheets one may obtain the symmetric solutions at
$-1<F_0<0$ (the line $AB$). They are placed on the unphysical sheets
($\tilde I$ and $\tilde I'$) that are  above the
physical sheet. In appendix A the construction of these sheets
is presented. In Fig.2 the horizontal shading marks the sheet
$\tilde I$. The dispersion equation (\ref{5}) on the unphysical
positive imaginary axis  has the same form on the sheets
$\tilde I$ and $\tilde I'$:
\begin{equation}\label{13}
1 + \frac 1{F_0} = -\gamma(arctg(\gamma) + \frac {\pi}{2}) .
\end{equation}
This equation turns into  Eq.(\ref{9}) if to change the sign of
$\gamma$. Solutions of this equation give the curve $AB$ in
Fig.2.

\begin{center}
{\bf III.C Solutions in RPA}
\end{center}

In Fig.3 the solutions of Eq.(\ref{1}) obtained in the kinetic
theory of Landau and in RPA are shown simultaneously
at $F_0$=-1.1, -1.2. The drawn solutions are  on the
upper semiplane of the physical sheet, i.e. they describe  the
unstable state of the nuclear matter.

The kinetic theory gives the linear dependence $\omega\sim
k$, that shows the unlimited undamping increasing of the
frequency with $k$. In RPA the dependence is different:
$Im\omega_{sd}(k)$ reach the maximum and then decrease. At
a certain value of $k$ (which we denote by $k_{fin}$)
 the branch $\omega_{sd}(k)$ goes over under the cut to
the unphysical sheet $I$. The values of the maximum
(which is proportional to the growth rate) and $k_{fin}$ depend
on parameters of matter $F_0$, $p_F$, $m$.

In Fig.4 the branches $\omega_{sd}(k)$ obtained in RPA are
presented. One sees that the branches change continuously  with
$F_0$ passing from the physical to unphysical sheets $I$.
At $-1 < F_0 < 0$  $\omega_{sd}(k)$ are placed on the unphysical
sheet completely.  At $F_0 < -1$ the branches lie on the
physical sheet at the momenta $k$ within the interval
$k=(0,k_{fin})$. All solutions belong to the same family
$\omega_{sd}$.

Similarly to Fig.2 the symmetrical branch
$-\omega_{sd}(k)$ can be found. At $k=k_{fin}$ it goes over from the
physical to unphysical sheets $\tilde I$.

There are two questions related to Fig.4. It is shown how the
branch go over the cut to unphysical sheet $I$ at $k =
k_{fin}$. But the cuts $I$, $I'$, $\tilde I$, $\tilde I'$ exist
when $k\leq 2p_F$.  If $\omega_{sd}(k)$ is on the unphysical
sheet at $k\approx 2p_F$, where will it be at $k > 2p_F$ when
the cut is closed and teared off the physical sheet?
One could show that $\omega_{sd}(k)$ keep on to exist on the
unphysical sheet.

The second question is what happens with $\omega_{sd}(k)$ when
$k_{fin}>2p_F?$  This takes place when $F_0 \ll -1$. Looking at
Fig.4 we wait that at $k = k_{fin}$ the imaginary branch
$\omega_{sd}(k)$ go over to unphysical sheet but $k_{fin}>2p_F$ and
the cut $I$ is closed. At $k_{fin}>2p_F$ there are two cuts
$III$ and $III'$ (\ref{12a}) on the real axis. They can be far
from the point of origin if $k_{fin}$ is distinctly larger
$2p_F$. It can be shown that two imaginary branches
$\omega_{sd}(k)$ and $-\omega_{sd}(k)$ turn to zero at $k =
k_{fin}$ and become real at $k > k_{fin}$. Further, as $k$
increase  they move along the real axis in the different sides,
reach the cuts $III$ and $III'$ and go over under the cuts. This
behaviour of branches is a standard one and is describe, for
example, in \cite{Tlr}.

\begin{center}
{\bf IV. Solutions at $F_0 > 0$}
\end{center}

The existence of zero-sound collective excitations at the
repulsive quasiparticle interaction  $F_0>0$ was predicted in
the paper \cite{La}.  The solutions of Eq.(\ref{1}) describing
the zero-sound excitations in kinetic theory are presented in
Fig.5.  They are the same as in Fig.1 in papers \cite{PR} and
\cite{CC}.

In RPA zero-sound excitations are described by the solutions of
Eq.(\ref{1}). Usually they are denoted by $\omega_s(k)$.
The excitations propagate with a small damping till the overlapping
of the  frequency $\omega_s(k)$  with the frequency of the free
particle-hole pairs. The overlapping takes place at a certain
wave vector $k$ that we denote $k_d$. It is $k=k_d$ when the
right point of the cut $II$ meets $\omega_s(k_d)$
\cite{PN}.  For $k>k_d$ the real solutions $\omega_s(k)$ that
correspond to the stable excitations, turn into the complex
ones. These complex solutions describe the damping excitations
and are situated on the unphysical sheet $II$, Fig.6.

The unphysical sheet $II$ is made by the same way as in the
previous section. The second logarithm in Eq.(\ref{10})
has the value $ln(-z_2)+\pi i$, where
$z_2= \frac{\omega m-kp_F-\frac12k^2}{\omega m-kp_F+
\frac12k^2}$, on the upper edge of the cut $II$
and   $ln(-z_2)-\pi i$ on the lower edge. We define as the
unphysical sheet $II$ the sheet neighboring with the physical
one where the magnitudes of $ln(z_2)$ differ by the quantity of $+2\pi
i$.  As before  going over from the physical sheet to the sheet $II$ we
have the continues changes of $ln (z_2)$ and, consequently, of
polarization operator.

In Fig.6 the sheet $II$ is marked by the
checked shading. In this figure $\omega_s(k)$ is presented in
the complex plane of $\omega$. At the wave vectors $k<k_d$,
the real branch $\omega_s(k)$ is  on the real axis.
At  $k=k_d$,  $\omega_s(k)$ becomes damping,  goes over
 the cut to the sheet $II$.

In Fig.7 the different models of $\omega_s(k)$  are
compared. The results are obtained at $F_0=2$, the corresponding
wave vector is $\frac{k_d}{p_F}=0.51$.
The solid line stands for the kinetic theory. The
dashed lines are for the real and imaginary parts of
$\omega_s(k)$ in RPA. The imaginary part of excitations appears due to
Landau damping \cite{LaLi}.

The Landau damping can be calculated by the
approximated equation \cite{FW}. Let us denote
$\omega_r=Re(\omega_s(k))$ and $\omega_i = Im(\omega_s(k))$.
Then at $\frac{\omega_i} {\omega_r}\ll 1$ the following
approximate equation is valid
\begin{equation}\label{14}
\omega_i = sign(\omega_r)Im\Pi^0(\omega_r,k)\left(\frac
{\partial Re(\Pi^0)}{\partial\omega}\right)_{\omega=\omega_r} .
\end{equation}
The $\omega_i$ obtained with this equation is shown by the
dotted line in Fig.7.

\begin{center}
{\bf V. Discussion}
\end{center}

The effective quasiparticle interaction used in this paper is
a simple constant interaction \cite{Mi}
\begin{equation}\label{d1}
{\cal F} = C_0(F + F'(\vec \tau_1\vec \tau_2)+ G(\vec
\sigma_1\vec \sigma_2) +  G'(\vec \sigma_1\vec
\sigma_2)(\vec \tau_1\vec \tau_2)),
\end{equation}
where $\vec \tau, \vec \sigma$ are the isospin and spin Pauli
matrices.  The results presented above are the valid for  all
types of the interaction in Eq.(\ref{d1}): scalar,
isospin, spin and spin-isospin one.

In the paper the unphysical sheets $I$ and $I'$ are defined. The
family of the branches of solutions $\omega_{sd}(k)$  is
placed (partly) on the imaginary unphysical axis where these
sheets coincide. It seems that the remaining surface of sheets
is not use.  But we can turn to the consideration of the pion
dispersion equation \cite{Mi} in the complex plane of $\omega$.
This equation is closely connected with the zero-sound
dispersion equation in the spin-isospin channel. It is appeared
that the solutions responsible for the pion condensation are
situated on the sheets $I$ and $I'$. When these solutions at a
certain conditions go over to the physical sheet (analogously
$\omega_{sd}(k)$ in Fig.4) then the physical values become
infinite. This is interpreted as a phase transition into another
state containing the pion condensate in the ground state
\cite{Mi},\cite{SR}.

The author thanks S.V. Tolokonnikov for the constructive critics
and  M.G.Ryskin for the fruitful discussions.
The work is supported by the RFBR (a grant number
03-02-17724).

\begin{center}
{\bf Appendix A}
\end{center}

Here we construct the unphysical sheets $\tilde I$ and $\tilde I'$.
The solutions of Eq.({5}) that are presented by  the curve
$AB$ in Fig.2 are placed on these sheets.

To pass on to the sheet $\tilde I$ we do the analytical continuation
in $\omega$ of
$ln(z)\equiv ln\left(\frac{\omega m}
{\omega m-kp_F +\frac{k^2}2}\right)$
from the lower edge of the cut $I$ upwards, adding $2\pi i$ to
$ln (z)$.
The sheet $\tilde I$ is situated above the physical sheet.
The value of $ln\left(\frac{\omega m}
{\omega m-kp_F +\frac{k^2}2}\right)$  above the upper edge of  the cut
$I$ is $ln(-z)-\pi i+2\pi i$. This expression is continued on the sheet
$\tilde I$.

In the similar way we construct the sheet $\tilde I'$ making the
analytical continuation of
$ln(z')\equiv ln\left(-\frac
{\omega m+kp_F -\frac{k^2}2}{\omega m}\right)$
from the lower edge of cut $I'$ adding $2\pi i$ to logarithm.

\newpage

\section{Figure captions}
\noindent
Fig.1. a)The complex plane of frequence.
The cuts of $\Pi^{0R}$  Eq.(\ref{3}) are shown. Numbers stand
for the cuts (\ref{12}).  b)The complex plane of $z_1$. The cut
corresponds to the cut $I$ in Fig.a. c)The complex plane of
frequence. The shading marks the unphysical sheets described in
the text.

\noindent
Fig.2. Solutions $\omega_{sd}(k)$ to Eq.(\ref{5}) at $F_0<0$.
The variable $\gamma$ is
$\gamma = \frac{m}{kp_F}Im~\omega_{sd}$.
The shading with the
right slope marks the unphysical sheet $I$.
The horizontal shading  marks the unphysical sheet $\tilde I$.

\noindent
Fig.3. The comparision of the solutions in the kinetic theory
and in the RPA. The solid (dashed) lines correspond to solutions
obtained at $F_0=-1.2$ ($F_0=-1.1$).

\noindent
Fig.4. The branches of solution $\omega_{sd}(k)$ obtained in
RPA at different values of $F_0$. The curve (1) corresponds to
$F_0=-0.4$;
(2) $F_0=-0.9$; (3) $F_0=-1.02$; (4) $F_0=-1.1$; (5) $F_0=-1.2$.
The shading marks the sheet $I$.

\noindent
Fig.5. The solutions to Eq. (\ref{5}) $\omega_s$ at  $F_0>0$.
Two symmetric solutions are presented.

\noindent
Fig.6. The complex plane of frequence. The solutions
$\omega_s(k)$ to Eq.(\ref{1}) obtained in RPA are presented. The
curve $1$ is calculated at $F_0=1$ and the curve $2$  at
$F_0=2$ . The wave vector $k_d$ marks the point when the Landau
damping starts at this $F_0$:
 $\frac{k_d}{p_F}=0.13$ when $F_0=1$,
and  $\frac{k_d}{p_F}=0.52$ at $F_0=2$.

\noindent
Fig.7. The branch $\omega_s(k)$ at $F_0=2$ is shown for the
different models. The solid line is for the solutions of
Eq.(\ref{5}). The real $\omega_r$ and the
imaginary $\omega_i$ parts of the RPA
solutions are presented by the dashed curves. The dotted line
stands for the $\omega_i$ calculated by Eq.(\ref{14}).

\newpage

\begin{figure}%1
%\vspace{-2cm}
\centering{\epsfig{figure=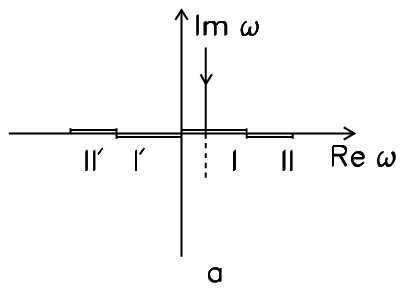,width=7cm}
\epsfig{figure=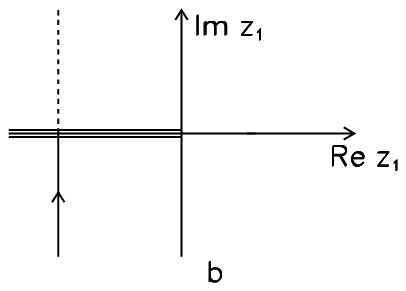,width=7cm}}

\vspace{1cm}

\centering{\epsfig{figure=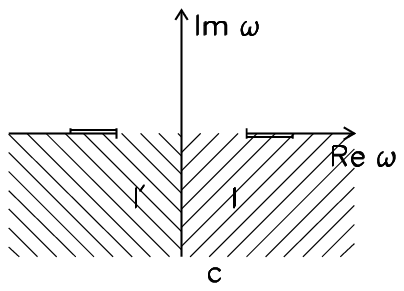,width=7cm}}
%\vspace{-1cm}
\caption{}
\end{figure}

\begin{figure}%2
\centering{\epsfig{figure=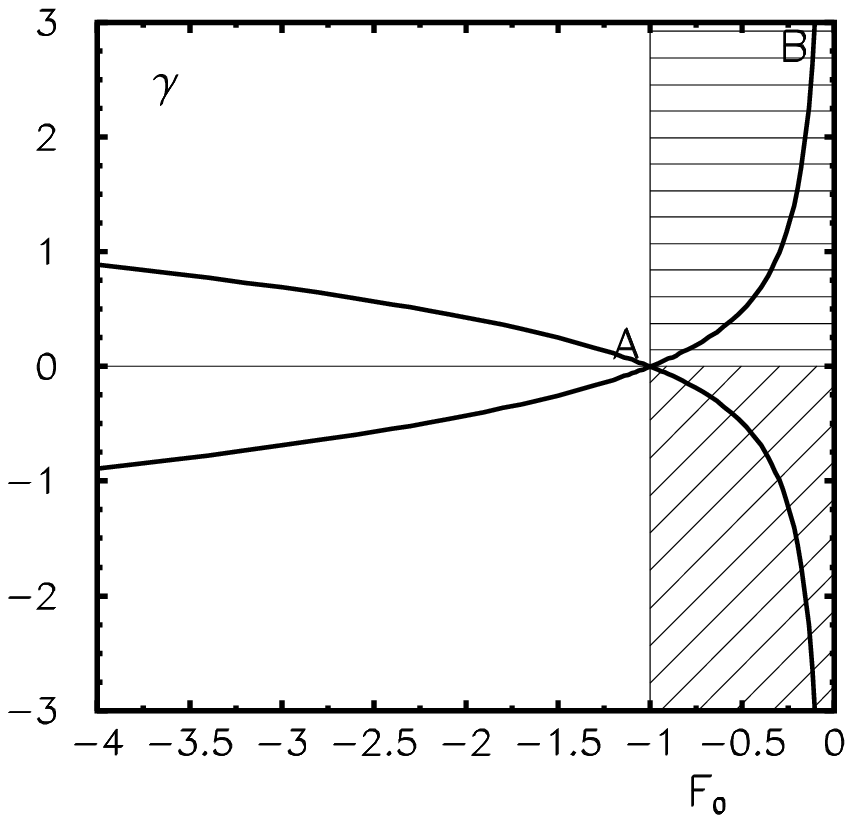,width=9cm}}
\caption{}
\end{figure}

\begin{figure}%3
\centering{\epsfig{figure=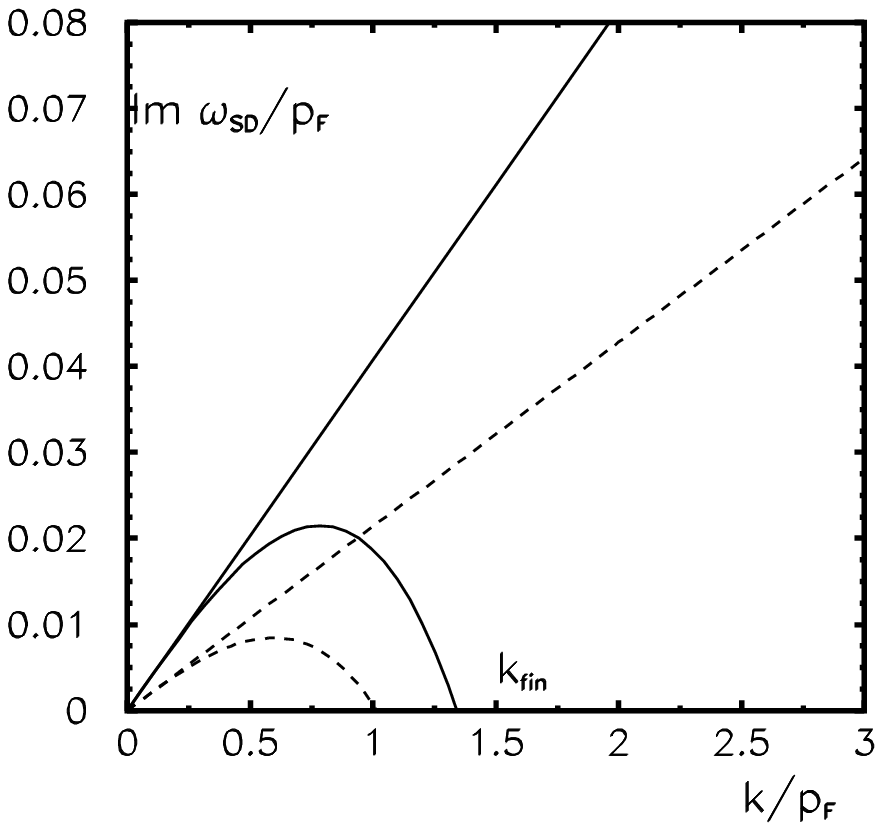,width=9cm}}
\caption{}
\end{figure}

\begin{figure}%4
\centering{\epsfig{figure=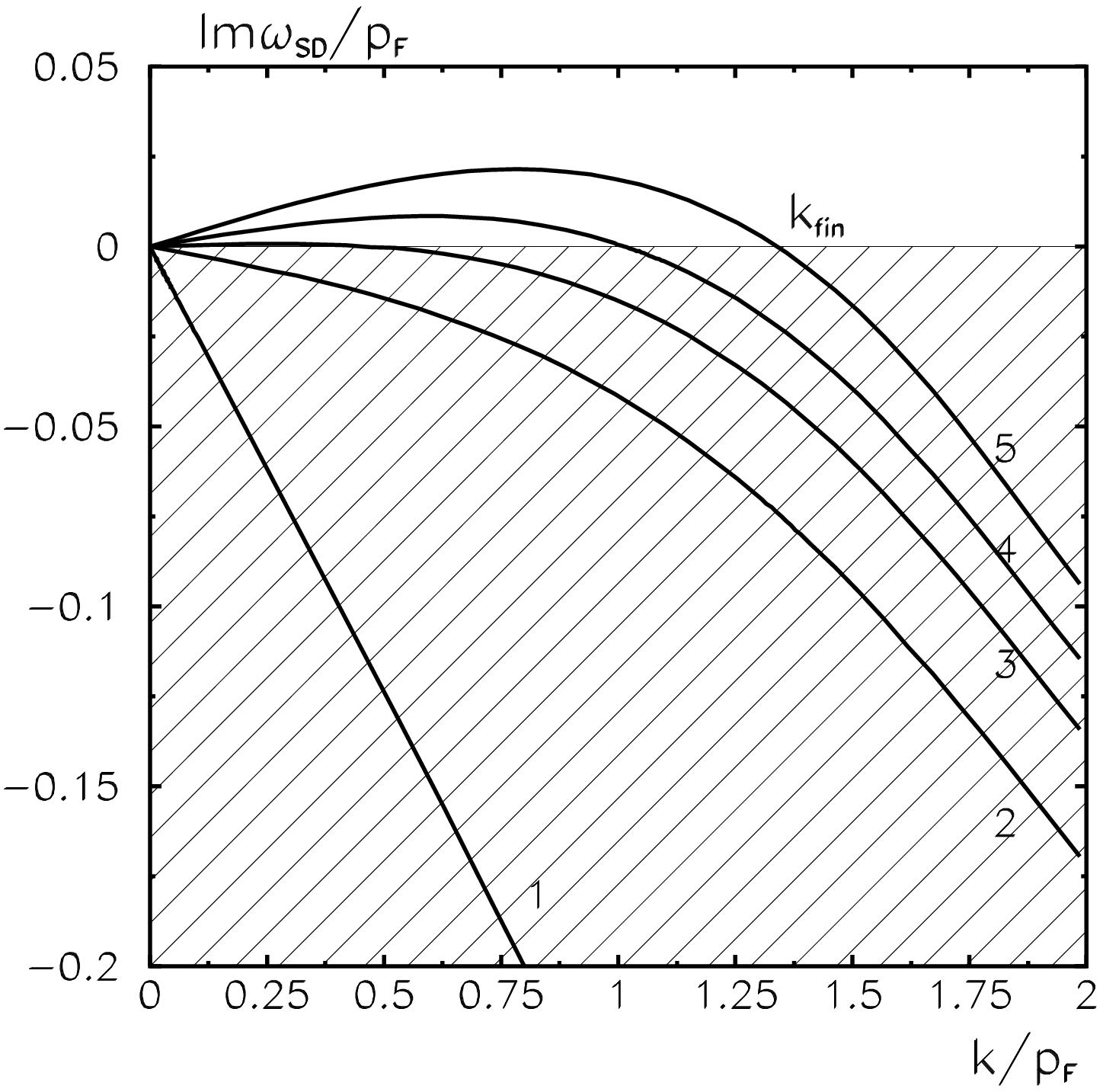,width=9cm}}
\caption{}
\end{figure}

\begin{figure}%5
\centering{\epsfig{figure=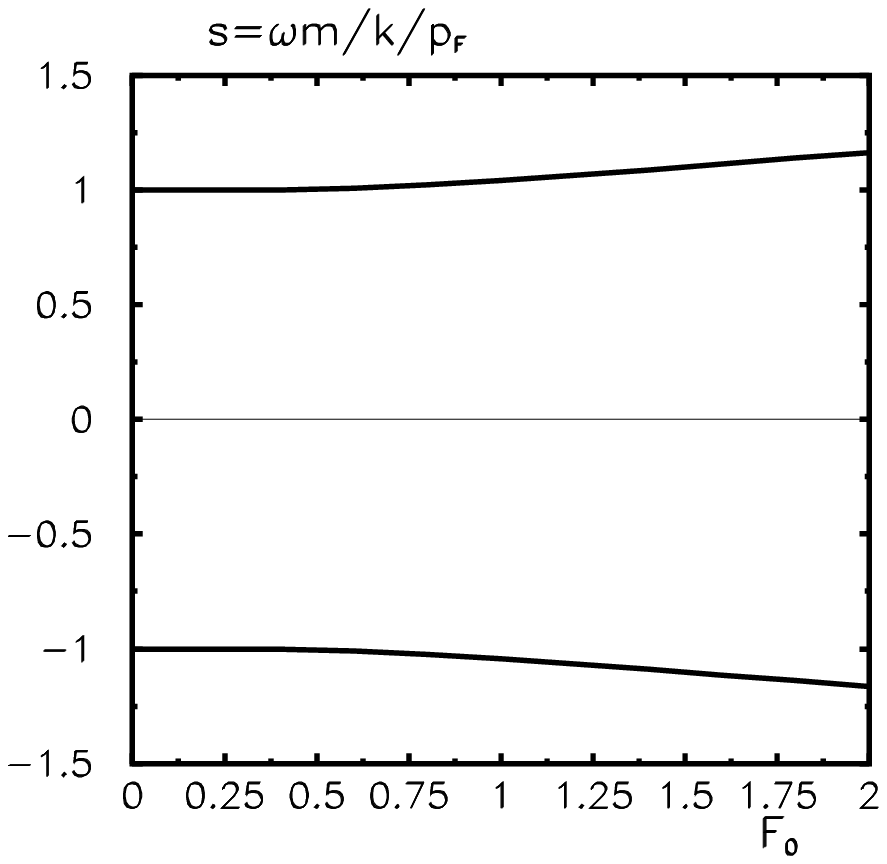,width=9cm}}
\caption{}
\end{figure}

\begin{figure}%6
\centering{\epsfig{figure=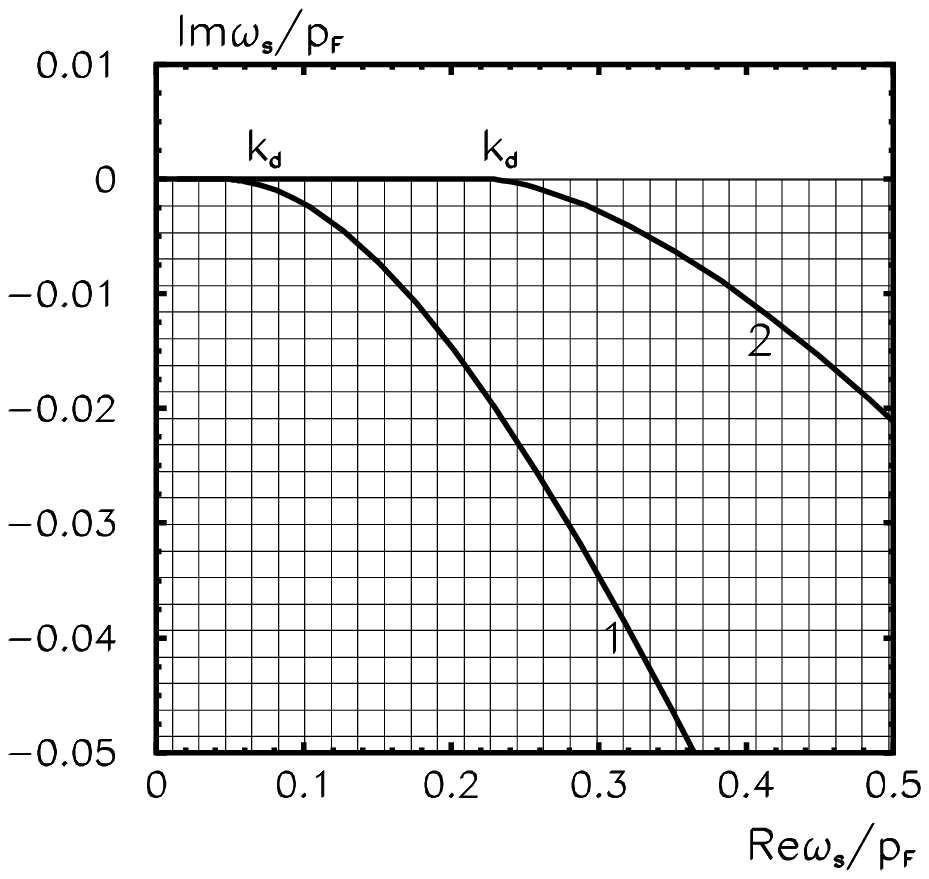,width=9cm}}
\caption{}
\end{figure}

\begin{figure}%7
\centering{\epsfig{figure=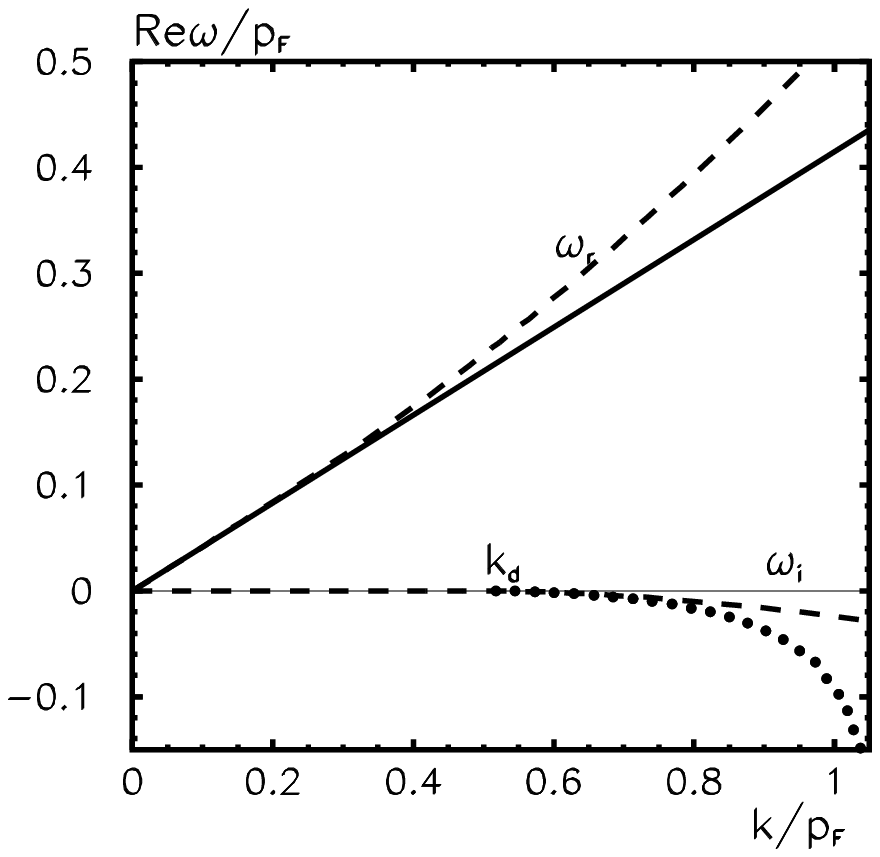,width=9cm}}
\caption{}
\end{figure}

\end{document}